%% file: main.tex
\documentclass[a4paper,gaps,prb,showpacs,twocolumn,superscriptaddress,footinbib]{revtex4}
\usepackage{amsmath,amssymb,mathrsfs}
\usepackage[utf8]{inputenc}
\usepackage[pdftex]{graphicx}
\usepackage[dvipsnames]{xcolor}
\usepackage[normalem]{ulem}
\usepackage[british]{babel}
\usepackage{csquotes}
\usepackage{comment}
\usepackage[colorlinks,citecolor=red,urlcolor=blue,bookmarks=false,hypertexnames=true]{hyperref}

\newcommand{\BW}{bandwidth\ }
\newcommand{\QUB}{School of Mathematics and Physics, Queen's University of Belfast, Belfast BT7 1NN, Northern Ireland, United Kingdom}
\newcommand{\ETSF}{European Theoretical Spectroscopy Facility (ETSF)}

\begin{document}

\title{A first-principles study and mesoscopic modeling of two-dimensional spin and orbital fluctuations in FeSe}

\author{Abyay Ghosh}
\author{Piotr Chudzinski} 
%\altaffiliation{E-mail: Piotr.Chudzinski@physik.uni-regensburg.de}
\address{\QUB}
\author{Myrta Gr\"uning}
\address{\QUB}
\address{\ETSF}
\date{\today}

\begin{abstract}

We calculated the structural, electronic and magnetic properties of FeSe within density-functional theory at the generalized gradient approximation level. First, we studied how the bandwidth of the $d$-bands at the Fermi energy are renormalized by adding simple corrections: Hubbard $U$, Hund’s $J$ and by introducing long-range magnetic orders. We found that introducing either a striped or a staggered dimer antiferromagnetic order brings the bandwidths---which are starkly overestimated at the generalized gradient approximation level---closer to those experimentally observed.
Second, for the ferromagnetic, the striped, checkerboard and the staggered dimer antiferromagnetic order, we investigate the change in magnetic formation energy with local magnetic moment of Fe at a pressure up to 6 GPa. The bilinear and biquadratic exchange energies are derived from the Heisenberg model and noncollinear first-principles calculations, respectively. We found a non-trivial behavior of the spin-exchange parameters on the magnetization, and we put forward a field-theory model that rationalizes these results in terms of two-dimensional spin and orbital fluctuations. The character of these fluctuations can be either that of a standard density wave or a topological vortex. Topological vortexes can result in mesoscopic magnetization structures. 
\end{abstract}

\maketitle

\section{Introduction}

The structurally simplest quasi-two-dimensional (2D) iron chalcogenide superconductor (ICS) FeSe \cite{doi:10.1073/pnas.0807325105} is of interest to the condensed matter and materials physics community because of the unconventional superconductivity, the not yet understood origin of the nematic phase and the absence of a long-range magnetic order~\cite{PhysRevLett.103.057002,cite-key,cite-key2,doi:10.1126/science.aal1575,doi:10.1073/pnas.1606562113}. The electronic structure and magnetic properties of the parent phase are intensely investigated, as they may help to explain those exotic properties. 

FeSe presents a peculiar electronic structure, that it shares with other ICSs. The $d_{xy}$ and $d_{xz/yz}$ orbital-derived bands around the Fermi level all contribute to the relevant physics of superconductivity, nematicity and magnetism, for which ICSs are said to have a multi-orbital multi-band nature. This makes the use of effective models more difficult than for cuprates, for which a single-band effective model is sufficient. On the other hand, standard density functional theory (DFT) calculations are known to strongly overestimate---by nearly a factor eight and four respectively~\cite{Yi2017}---the bandwidth of the relevant bands around the Fermi level when compared with ARPES experiments.~\cite{PhysRevB.91.155106,Yi2017,PhysRevX.9.041049}  

The renormalization of the conventional DFT bandwidth is the signature of correlation, which is missing at this level of theory. 
Calculations including strong local correlation at different levels of theory~\cite{PhysRevB.104.L161110,LOHANI201554,PhysRevLett.119.067004,sym13020169,cite-key7a,PhysRevLett.121.197001}  (addition of Hubbard's $U$, slave-boson theory, Density Mean Field Theory either on top of conventional DFT, hybrid DFT or quasiparticle GW) have been performed with partial success in reproducing the bandwidth of the relevant bands. A different approach has been taken in Ref.~\cite{PhysRevB.102.235121} which applied conventional DFT to a paramagnetic supercell and found the bandwidth of the $d_{xy}$ and $d_{xz/yz}$ bands is strongly renormalized compared to nonmagnetic calculations, in quite good agreement with the ARPES results. Further, their effective band structure also reproduces the large broadening of the  $d_{xy}$ band around the $\Gamma$ point. The latter work points to a strong interplay between electronic and magnetic properties and the need to consider the magnetic structure of FeSe to reproduce its electronic structure.       
Indeed, despite the absence of long-range magnetic order at ambient pressure, the consensus is that FeSe serves as a platform for diverse competing magnetic interactions such as N\'eel, stripe, or staggered-type antiferromagnetic interactions \cite{Glasbrenner2015,Wang2016,cite-key6,Fernandes2022}. Applying hydrostatic pressure around 2 GPa induces stripe-type long-range order into the system \cite{Sun2016}.  
Besides inducing long-range magnetic order into the system, the pressure-temperature phase diagram also reveals that pressure suppresses nematicity while superconducting T$_c$ is enhanced by factor four~\cite{Sun2016,cite-key4}. Both spin and orbital degrees of freedom are supposed to play a key role in this phase diagram \cite{PhysRevX.10.011034}.

In this work, using DFT, we investigate the coupling of spin and orbital degrees of freedom, thus the interplay of electronic structure and magnetic properties of FeSe in its parent phase. Using field theory, we then consider the implications of such coupling, seeking to 'capture the missing correlation' in standard DFT nonmagnetic calculations. This investigation is articulated into three parts.   

 In the \emph{first} part (Sec.\ref{sec:DFTresult}), we systematically study how the electronic structure from conventional DFT, and specifically the bandwidths of the $d_{xy}$ and $d_{xz/yz}$ orbital-derived bands, is modified when adding on-site magnetization, Hubbard $U$ and Hund's $J$. Remarkably, we find that a reasonable comparison with experimental data is obtained for phases hosting a magnetic order, either SAFM or staggered dimer. This is in agreement with a finite on-site magnetization that minimizes the DFT energy. However, experiments do not support the existence of a macroscopic spin order at ambient pressure. 

In the \emph{second} part (Sec.\ref{sec:DFTmagnetic}), we compute within DFT the magnetic formation energy as a function of the local magnetic moment for varying pressure, so probe the energy landscape of different magnetic interactions. Since the local magnetic moment affects orbital occupation, these calculations provide insight into the role of the orbital degree of freedom, as highlighted in Refs. ~\cite{PhysRevB.104.L241101,cite-key6a}. We map these results onto a Heisenberg-like Hamiltonian~\cite{PhysRevB.89.064509,cite-key7,PhysRevB.106.L060504}
to find a non-trivial dependence of the spin-exchange parameters on the local magnetic moments, which points to a strong coupling of the spin and orbital degrees of freedom.

The implications of this strong coupling are considered in the \emph{third} part (Sec.\ref{sec:QFTresult}) at the many-body level, thus including collective phenomena, using field-theory modeling. We formulate a theory which describes the spin and orbital collective fluctuations as two coupled 2D classical degrees of freedom, described by Kosteritz-Thouless (KT) type theories. Our model can capture highly nonlocal features and in particular can host mesoscopic patterns of orbital configurations, i.e. vortexes in KT language. The motivation of our study, looking towards mesoscopic domains/patterns, comes also from recent experimental reports where nanoscopic Griffiths phases were found in FeSe \cite{PhysRevLett.127.246402}.  
%The experimental implications of this approach are summarized in section.

\section{Computational Details}\label{sec:method}
The plane wave pseudopotential suite QUANTUM ESPRESSO~\cite{Giannozzi_2009,Giannozzi_2017} is used to perform fully self-consistent DFT-based electronic structure calculations by solving the standard Kohn-Sham (KS) equations. Ultrasoft pseudopotentials from the PSlibrary~\cite{DALCORSO2014337} are used for Fe and Se atoms. Kinetic-energy cut-offs are fixed to 55 Ry for electronic wave functions after performing rigorous convergence tests.

The electronic exchange-correlation is treated under the generalized gradient approximation (GGA) that is parametrized by Perdew-Burke-Enzerhof (PBE) functional~\cite{PhysRevLett.77.3865,PhysRevLett.78.1396}. 
Hubbard's parameters used in the DFT+$U$ calculations are determined from a piecewise linearity condition implemented through linear-response theory~\cite{PhysRevB.71.035105}, based on Density Functional Perturbation Theory (DFPT)~\cite{PhysRevLett.58.1861} as implemented in QUANTUM ESPRESSO. A dense $q$-mesh grid of 3$\times$3$\times$3 is considered for the DFPT calculation. We  have obtained a Hubbard $U$ of 6.90 eV. 
 
Supercell of size 2$\times$2$\times$1 is used to calculate magnetic moment-dependent energies in different magnetic phases like ferromagnetic (FM), checkerboard antiferromagnetic (CAFM), striped antiferromagnetic (SAFM) and staggered dimer (SD). We adopt the Monkhorst-Pack scheme~\cite{PhysRevB.13.5188} to sample the Brillouin zone in k-space with 8$\times$8$\times$8 grid.  Band unfolding technique as implemented in BandUPpy module was used to get primitive cell band structure from supercell magnetic lattice~\cite{IRAOLA2022108226,PhysRevB.89.041407,PhysRevB.91.041116}. 

The coordinates are optimized for each magnetic phase. Geometry optimization has been performed using the Broyden-Fletcher-Goldfrab-Shanno (BFGS) scheme \cite{BFGS_Opt}. The experimental lattice parameters ($a,b$ = 3.7698 $\AA$, $c =$$ 5.5163 \AA$ and $z_{Se}$ = 0.2576) are used as starting values. The Convergence threshold of $10^{-8}$ and $10^{-3}$ are used on total energy (a.u) and forces (a.u) respectively for ionic minimization. High-pressure structures are obtained by enthalpy ($H=U+PV$) minimization under externally applied hydrostatic pressure. Fixed volume coordinate optimization calculation has been performed with long-range magnetic order at all pressures. 

Energy penalty functional is used to perform the constrained magnetic moment calculations. The penalty term is incorporated into total energy by weight $\lambda$ as: $E_{total}=E_{LSDA}+\sum_{i}\lambda(M_{i}-M^0_{i})^2$, where $i$ is the atomic index for Fe atoms and $M^0_{i}$ is the targeted local magnetic moment at atom $i$. The value of $\lambda$ is fixed to 25$Ry/\mu^2_{B}$ after performing a convergence test, constraining the magnetic moment of Fe at a particular value. The angular dependence of energy is calculated by performing fully noncollinear first-principles calculations.

\section{Bandwidth renormalization}\label{sec:DFTresult}

We present the results for the three $t_{2g}$ orbital-derived bands, $d_{xy}$, $d_{xz}$, $d_{xy}$ which are dominant to the hole-like bands crossing the Fermi level around $\Gamma$ point. We employ DFT (without and with spin polarization), DFT+$U$, DFT+$J$ both in the nonmagnetic (NM) case and assuming striped antiferromagnetic (SAFM) or staggered dimer (SD) order. We focus on the band dispersion, along the $\Gamma-M$ direction, and consider the difference between the maximum and the minimum band energy which in what follows we refer to simply as bandwidth. It has been observed from ARPES experiments, that conventional nonmagnetic DFT strongly overestimates the bandwidth of both $d_{xy}$ and $d_{xz/yz}$ bands. Here, we observe the effect of the on-site Hubbard repulsion, of Hund's $J$, which is predicted to play an important role for FeSe compounds~\cite{PhysRevX.10.011034,sym13020169}, and SAFM and SD order. Though FeSe does not present a long-range magnetic order---contrary to ICSs which mostly order magnetically---the SAFM has been observed to be the dominant magnetic fluctuation in FeSe \cite{Wang2016} and SAFM order appears in the system at higher pressures \cite{Matsuura2017}. We also considered the SD phase which has been predicted theoretically to be slightly lower in energy than SAFM.~\cite{PhysRevB.93.205154}   Results of the \BW of the $t_{2g}$ orbital-derived bands are summarised in Table-\ref{bw}.          
%------------------------------------------------------------------------------------------------------------------------------------
  \begin{table}[!htbp]
  \tiny
  \caption{Approximate values of bandwidth (in meV) along the $\Gamma-M$ direction of the $d_{xy}$, $d_{xz}$, $d_{xy}$ orbital derived bands at different levels of the theory and for different magnetic phases.}
  \label{bw}
  \resizebox{\columnwidth}{!}{
  \begin{tabular}{|lc|c|c|c|c|c|}
  \hline
  \multicolumn{6}{|c|}{Non Magnetic GGA/GGA+U} \\ 
  \hline
  &\text{Orbital}           & \text{GGA}     & \text{$U$ = 3.45 eV}       & \text{$U$ = 6.90 eV}        & \text{-} \\ 
  \hline
 & \text{$d_{xy}$}               &  834.8               & 698.5               &  565.8                  &  - \\
  & \text{$d_{xz/yz}$}               &  572.8               & 632.2               & 681.1                 &  - \\
   \hline
   \multicolumn{6}{|c|}{Spin polarized GGA/GGA+U} \\ 
    \hline
    &\text{Orbital}           & \text{GGA}     & \text{$U$ = 3.45 eV}       & \text{$U$ = 6.90 eV}         & \text{-}  \\ 
  \hline
  & \text{$d_{xy}$}              & 645.1               & 579.8               & 160.7         & \text{-}  \\
  & \text{$d_{xz/yz}$}               & 436.5               &  468.0               & 995.3          & \text{-}  \\
  \hline 
   \multicolumn{6}{|c|}{Spin polarized GGA+J} \\ 
    \hline
    &\text{Orbital}           & \text{$J$ = 0.05 eV}     & \text{$J$ = 0.10 eV}       & \text{$J$ = 0.20 eV}       & \text{$J$ = 0.35 eV}  \\ 
  \hline
  & \text{$d_{xy}$}              & 621.7               & 600.6               & 555.4               & 454.0  \\
  & \text{$d_{xz/yz}$}               & 429.5               &  426.0               & 415.5               & 398.1   \\
  \hline    
     \multicolumn{6}{|c|}{Spin polarized with long-range magnetic order} \\ 
    \hline
    &\text{Orbital}           & \text{SAFM}     & \text{$U$ = 1.0 eV}       & \text{$U$ = 2.0 eV}       & \text{$U$ = 3.4 eV}  \\ 
  \hline
  & \text{$d_{xy}$}              & 227.8               & 202.9               & 238.9               & 430.2  \\
  & \text{$d_{xz/yz}$}               & 271.1               &  476.4               & 551.7               & 636.1   \\
  \hline 
     &\text{Orbital}           & \text{$J$ = 0.05 eV}     & \text{$J$ = 0.10 eV}       & \text{$J$ = 0.20 eV}       & \text{$J$ = 0.50 eV}  \\ 
  \hline
  & \text{$d_{xy}$}              & 272.1              & 303.8               & 447.8               & 712.9  \\
  & \text{$d_{xz/yz}$}               & 256.7               &  274.2               & 274.6               & 356.4   \\
  \hline  
  \end{tabular}
    }
  \end{table}
%------------------------------------------------------------------------------------------------------------------------------------

%Table \ref{bw} shows the calculated values of the $d_{xy}$, $d_{xz}$, $d_{yz}$ \BW.  %which are dominant to the hole-like bands crossing the Fermi level around $\Gamma$ point. A ladder approach has been incorporated to improve the \BW values. 

\textit{NM GGA and GGA+U}. The values calculated at the NM GGA level, 834.8 meV, and 572.8 meV for $d_{xy}$  and $d_{xz/yz}$ respectively, are in agreement with \cite{Yi2017}.  As already reported~\cite{PhysRevB.91.155106}, the \BW of these bands are strongly overestimated compared to experiments. 
Next, we add Hubbard $U$ correlation, using for $U$ both the value determined from DFPT and half of such value, to study the dependence of the \BW on $U$.  The determined U value for FeSe within DFPT stands at 6.90 eV, surpassing the 4.06 eV derived from constrained random phase approximation. \cite{PhysRevB.82.064504,PhysRevB.95.081106}.
The effect of $U$ is strongly orbital dependent: by increasing $U$, the \BW of $d_{xy}$ decreases whereas the of $d_{xz/yz}$ increases compared to NM GGA.
The simple correction using a mean-field $U$ is insufficient here due to the multi-orbital, multi-band nature of the system.

\textit{Spin polarized GGA and GGA+U}. By just considering spin polarization the \BW of NM GGA is renormalised by about a factor 1.3. As in the NM case, $U$ is orbital selective. The \BW of the $d_{xy}$ derived band is renormalised to 160.7 meV at $U$ = 6.90 eV, while the \BW of $d_{xz/yz}$ increases to the value of 995.3 meV. 

\textit{Spin polarized GGA+J}: Adding increasing Hund's $J$ show a renormalization of \BW of the $t_{2g}$ orbital-derived bands. At $J$ = 0.35 eV, the \BW of $d_{xy}$ and $d_{xz/yz}$ is reduced to 454.0 meV and 398.1 meV respectively. However, when further increasing $J$ the nature of the bands changes bringing the calculated bandstructure in qualitative disagreement with the experimental observations. The combined addition of  Hubbard $U$ and Hund's $J$ (results not shown) does not bring any improvement in applying the correction separately.

\textit{Long range magnetic order GGA, GGA+U, GGA+J}:  The \BW of both $d_{xy}$ and $d_{xz/yz}$ orbitals are renormalised to 272.1 meV and 256.7 meV respectively in SAFM phase. As previously observed, the addition of $U$ is orbital selective, at least in the case of "small" $U$. For $U$=1~eV, the \BW of $d_{xy}$ decreases to 202.9 eV and the $d_{xz/yz}$ \BW increases to 476.4 eV. Instead, larger values of $U$ in combination with magnetic ordering cause an increase of the \BW of both the orbitals. 
The addition of a small Hund's $J$ (0.05 eV) increases the \BW of $d_{xy}$ slightly to 271.1 meV and decreases that of $d_{xz/yz}$ to 256.7 meV. As for the Hubbard parameters, larger values of $J$ in combination with magnetic ordering cause an increase of the \BW of both the orbitals. These results indicate that considering the magnetic ordering, also accounts for most of the effects of adding the Hubbard $U$ and Hund's $J$. 

  \begin{table}[!htbp]
  \tiny
  \caption{Comparison of bandwidth (in meV) from different methods}
  \label{bw1}
  \resizebox{\columnwidth}{!}{
  \begin{tabular}{|lc|c|c|c|c|c|}
  \hline
  &\text{Orbital}           & \text{DFT+DMFT}     & \text{DFT@SAFM}     & \text{DFT@SD}       & \text{Expt.}       \\ 
  \hline
  & \text{$d_{xy}$}              & 225.0              & 227.8           & 169.8            & 37.5               \\
  & \text{$d_{xz/yz}$}               & 200.0               &  271.1     & 367.8           & 155.0            \\
  \hline  
  \end{tabular}
    }
  \end{table}

%------------------------------------------------------------------------------------------------------------------------------------
%Another important magnetic phase predicted from theoretical calculations is SD which is slightly lower in energy than SAFM~\cite{PhysRevB.93.205154}. 
Table-\ref{bw1} compares the \BW we calculated within GGA in the SAFM and SD long-range magnetic order, with those obtained from DFT+DMFT \cite{PhysRevLett.119.067004} and experimentally observed values. The \BW of $d_{xy}$ obtained from DFT+DMFT is 225.0 meV, very close to the value 227.8 meV we obtained assuming a SAFM magnetic order. The \BW is further reduced to 169.8 meV by assuming the SD magnetic order. All these values are still too large by a factor 4--6 compared with the value of 37.5 meV extracted from experiments.~\cite{PhysRevB.91.155106}
The calculated bandwidth of the $d_{xz/yz}$ band is 200.0 meV, 271.1 meV, and 367.8 meV for DFT+DMFT and assuming a SAFM and SD magnetic order respectively. This is closer to the value extracted from the experimental which is about 155.0 meV. Then, assuming a long-range magnetic order has a similar effect as introducing strong electron correlation through DMFT. Both DFT+DMFT and DFT with a long-range magnetic order predict the \BW of the $d_{xy}$ and $d_{xz/yz}$ orbital-derived bands to be of the same order and so overestimating by a factor 4--6 larger the \BW of the $d_{xy}$ band.
    
\section{Magnetic properties}\label{sec:DFTmagnetic}

\begin{figure}[h!]
 \centering
 \includegraphics [width=0.46\textwidth]{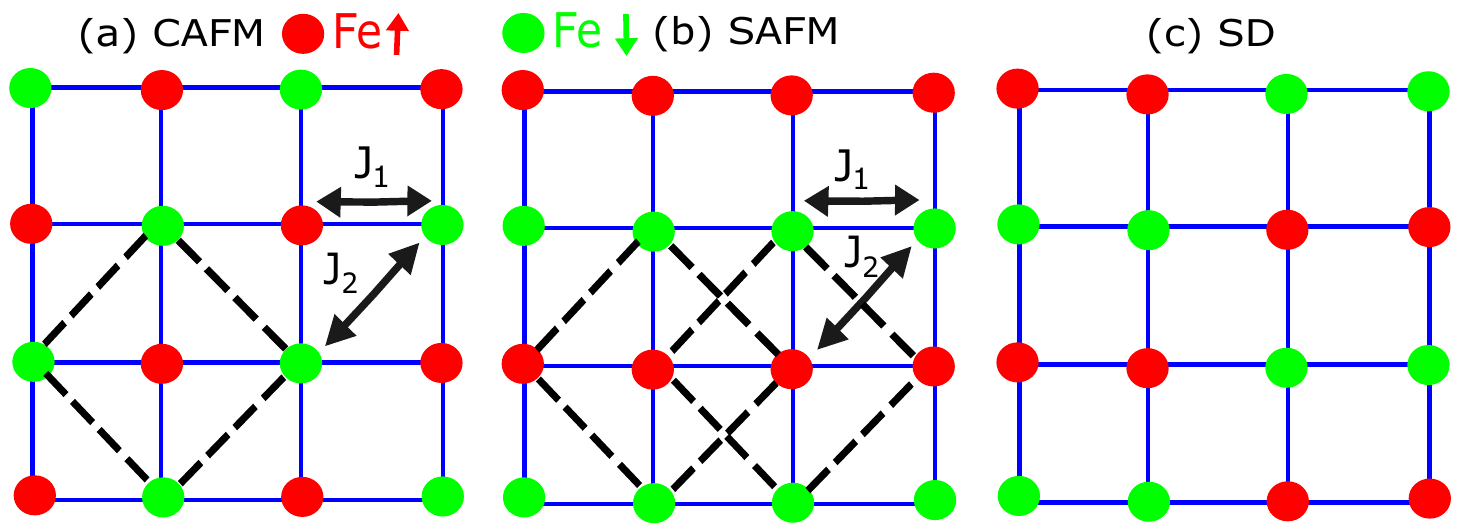}
 \caption{Spin arrangement in FeSe magnetic lattice: (a) checkerboard antiferromagnetic (CAFM), (b) striped antiferromagnetic (SAFM), and (c) staggered dimer (SD). The red and green circles represent Fe atoms in the lattice with up and down spin respectively. The black dashed lines highlight the  AFM square (sub)lattice. The SAFM phase can be imagined as two interpenetrating AFM square sublattices. $J_1$ and $J_2$ are the nearest neighbors (NN) and next nearest neighbor (NNN) spin-exchange parameters in the Heisenberg model (see Sec.~\ref{ss:Heisenberg}).}
 \label{mag}
\end{figure}

We consider four different magnetic phases: ferromagnetic (FM), checkerboard antiferromagnetic (CAFM), striped antiferromagnetic (SAFM) and staggered dimer (SD). The three antiferromagnetic  (AFM) configurations are shown in Fig.\ref{mag}(a,b,c). In the checkerboard phase [Fig.\ref{mag}(a)], the nearest neighbor (NN) spins are anti-parallel to each other. In the striped phase [Fig.\ref{mag}(b)], sites with the same spin form a stripe between stripes of opposite spin so that the next nearest neighbor (NNN) spins are antiparallel. In the staggered dimer phase, as the name suggests, sites with the same spin form dimers and (NN) dimers have opposite spins.  
For each phase, at different hydrostatic pressures, we calculate the dependence of magnetic formation energy on the local magnetic moment (part~\ref{ssec:magneticp}). With such calculations, one can probe the magnetic phase in a high-spin or low-spin state and identify possible metastable states within the subspace of a given constrained magnetic moment. Then, we map these results into a Heisenberg model (part~\ref{ss:Heisenberg}) and study the dependence of the spin-exchange parameters on the local magnetic moment at different hydrostatic pressures. Since changing the local magnetic moment corresponds to changing the orbital configuration, these calculations allow us to investigate the coupling of the orbital and spin degrees of freedom. Finally, we inspect the density of states (DOS) close to the Fermi energy (part~\ref{ssec:HS-LS}). 

%HERE? The SAFM phase can be imagined as two interpenetrating AFM square sublattices. Each sublattice can be rotated with respect to the other without costing any energy. Hence, the SAFM configuration is fully frustrated. This kind of frustrated magnetism is found to be associated with structural distortions and quantum fluctuations.~\cite{PhysRevLett.101.057010}  ←this is for LaFeAsO, 
%------------------------------------------------------------
   %-------------------------------------------------------
\subsection{Magnetic formation energy dependence on magnetization}\label{ssec:magneticp}
\begin{figure}
 \centering
 \includegraphics [width=0.40\textwidth]{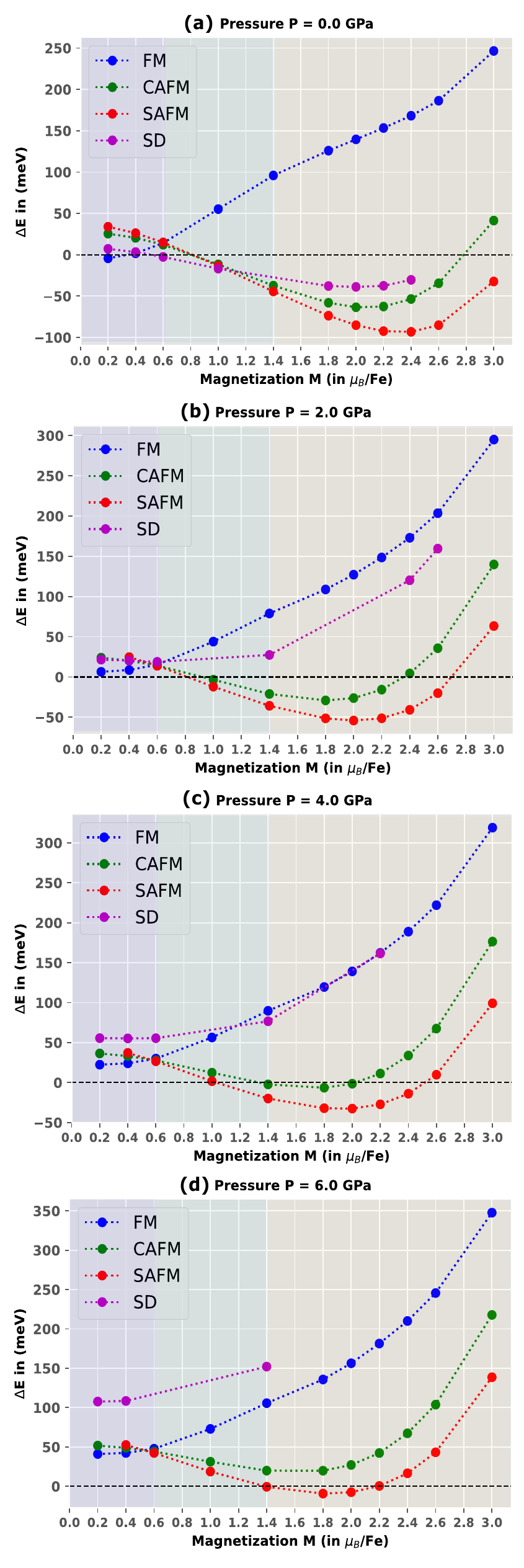}
 \caption{ Dependence of the magnetic formation energy with the local magnetic moment in (a) at ambient pressure (P), and with pressure (b) $P$ = 2.0 GPa, (c) $P$ = 4.0 GPa, (d) $P$ = 6.0 GPa.}
 \label{M}
\end{figure}

The magnetic formation energy, $\Delta E$, of a magnetic phase is defined as the energy difference per atom between the system in the magnetic phase (at a magnetic moment $M$) and the non-magnetic phase. A magnetic phase is energetically favourable when $\Delta E<0$. 
In Fig.~\ref{M}, we plot the magnetic formation energy, $\Delta E$, against the magnetic moment $M$ at different values of the applied pressure for the considered magnetic phases.  In the following discussion, we distinguish a low ( $0.2\mu_B\leq M \leq 0.6\mu_B$), intermediate ( $0.6\mu_B\leq M \leq 1.4\mu_B$ ) and high ( $1.4\mu_B\leq M \leq 3.0\mu_B$) magnetization region. Correspondingly to these regions, we also refer to low-spin (LS) and high-spin (HS) states as low values of Fe local magnetization correspond to a low-spin state configuration of the Fe atom, while high values of Fe local magnetization correspond to a high-spin state configuration.  
%------------------------------------------------------------
   %----------------------------------------------------------

At ambient pressure (Fig.\ref{M}(a)), in the LS state, the FM and the SD phases are the competing stable phases. At  $M = 0.2 \mu_B$, the FM phase is the only (slightly) energetically favourable phase ($\Delta E =-$4.32 meV). As the local magnetic moment is increased to 0.6$\mu_B$, a magnetic transition from FM to SD occurs and the SD phase becomes stable, while the FM becomes unstable. Between 1.0–1.2$\mu_B$, the other AFM phases become stable and the three AFM phases are nearly degenerate. In the HS state, the AFM are stable (up to 2.8 $\mu_B$) and reach their minimum. 
%This contrasts with most iron pnictide superconductors where only stripe-type AFM order is observed to be stable.~\cite{Fernandes2022} 
The energy minima for SD ($\Delta E = -$45.72 meV) and CAFM ($\Delta E = -$ 63.5 meV) are found for a magnetization around 2.0$\mu_B$. SAFM is the most stable with $\Delta E = -$93.2 meV around a magnetization of 2.3$\mu_B$. As these energies are much larger in absolute value than the transition temperature (or even room temperature), these calculations predict that these magnetic phases should be thermally stable. In contrast, the experimental phase diagram of FeSe at ambient pressure shows no long-range magnetic order phases although fluctuations---both of SAFM at ($\pi$,0)  and CAFM  at ($\pi$,$\pi$)---have been observed in neutron scattering measurements over a wide energy range \cite{Wang2016}. Also, the observed magnetic moment in the experiment is 2.28$\mu_B$ which is close to the optimal magnetic moment of Fe we predict in the SAFM phase.

At a pressure $P =$ 2.0 GPa [Fig.\ref{M}(b)], the FM and the SD are unstable at all values of the magnetization. Thus, no stable phases are observed at low spin (though we did not include spin-orbit coupling which can potentially stabilize the FM or SD phase).
The magnetic formation energy minima of the CAFM and SAFM are reduced in absolute value ($\Delta E = -$29.2 meV and $\Delta E = -$54.1 meV respectively). According to these calculations then, the SAFM phase should be stable, in agreement with what observed experimentally~\cite{Sun2016} at low temperatures. Also, the CAFM fluctuations should thus be still observable at this pressure. Further, the reduced energy difference CAFM-SAFM difference can allow spin-flip processes between these two phases.

As the pressure is increased to 4.0 GPa [Fig.\ref{M}(c)], the absolute values of the magnetization formation energy of CAFM and SAFM  are reduced further to $\Delta E = -$ 6.4 meV and $\Delta E = -$32.6 meV.  At pressure 6.0 GPa, the CAFM phase is unstable throughout the range of magnetic moment [Fig.\ref{M}(d)]. The majority of the striped fluctuations are eliminated and the energy minimum is shifted to $\Delta E = -$-9.3 meV. Consequently, in agreement with experiments long-range magnetic order disappears at about 6.0 GPa.

Table.\ref{mm} shows how the optimal value of the magnetic moment in CAFM and SAFM phase decreases when increasing the pressure. 
%---------------------------------------------------------
  \begin{table}[!htbp]
  \tiny
  \tabcolsep=0.3cm
  \caption{Calculated optimal local magnetic moment of Fe (in $\mu_B$) for the checkerboard (CAFM) and striped (SAFM) antiferromagnetic phases for increased applied hydrostatic pressure. Here, optimal magnetization is the magnetization corresponding to the minimum value of the magnetization formation energy for the given phase. For increasing pressure, the optimal local magnetic moment of Fe decreases for both phases.}
  \label{mm}
  \resizebox{\columnwidth}{!}{
  \begin{tabular}{|lc|c|c|c|}
  \hline
  &\text{Pressure (GPa)}           & \text{CAFM}     & \text{SAFM} \\ 
  \hline
 & \text{0.0}               &  2.08               & 2.32    \\
  & \text{2.0}               & 1.78               & 1.99   \\
  & \text{4.0}               &  1.70               & 1.91    \\
  & \text{6.0}               & 1.60               & 1.83   \\
  \hline  
  \end{tabular}
    }
  \end{table}

%------------------------------------------------------------------------------------------------------------------------------------
%The above discussion clearly shows that FeSe at ambient pressure is the host of several competing magnetic orders in magnetic moment space. At the minimum of magnetization (high spin state), their energy difference is around 40meV at $p=0$ and 20meV at higher pressures, with a notable exception of the SD phase that continuously moves to higher energies with pressure. The long-range magnetic order is eventually removed because it becomes energetically unstable at higher pressures. 

%\subsection{Nearest neighbor ($J_1$), next nearest neighbor ($J_2$) and bi-quadratic (K) interaction}\label{ss:Heisenberg}
\subsection{Heisenberg model Hamiltonian}\label{ss:Heisenberg}
We map the total energies for the considered phases into a Heisenberg-like model including nonlinear terms:
%------------------------------------------------------------
\begin{equation}\label{1}
\small
H=J_1\sum_{ij=NN}\vec{S}_i\cdot\vec{S}_j + J_2\sum_{ij=NNN}\vec{S}_i\cdot\vec{S}_j – K\sum_{ij=NN}(\vec{S}_i\cdot\vec{S}_j)^2,
\end{equation} 
%------------------------------------------------------------
where $J_1$, $J_2$ and $K$ represent NN, NNN and bi-quadratic exchange interaction parameters respectively. $\vec{S}_i$, $\vec{S}_j$ are spin magnetic moment at site $i$ and $j$ respectively. We choose not to include the third nearest neighbor term $J_3$ since it was found to be significantly smaller than $J_1$($J_2$ or $K$) for FeSe~\cite{PhysRevB.106.L060504}. AFM (FM) states are defined by positive (negative) $J_1$, $J_2$. 

The bi-quadratic term, $K$, is calculated from a series of non-collinear calculations performed by varying the angle, $\theta$, between two magnetic sublattices~\cite{Glasbrenner2015,PhysRevB.89.064509} in the SAFM phase (see Fig.~\ref{mag}). The $K$ parameter is then extracted by fitting the angular energy dependence $E(\theta)$ with 
%------------------------------------------------------------
\begin{equation}\label{4a}
E(\theta) - E(0) = 2K\sin^2\theta.
\end{equation} 
%------------------------------------------------------------

\begin{figure}
 \centering
 \includegraphics [width=0.425\textwidth]{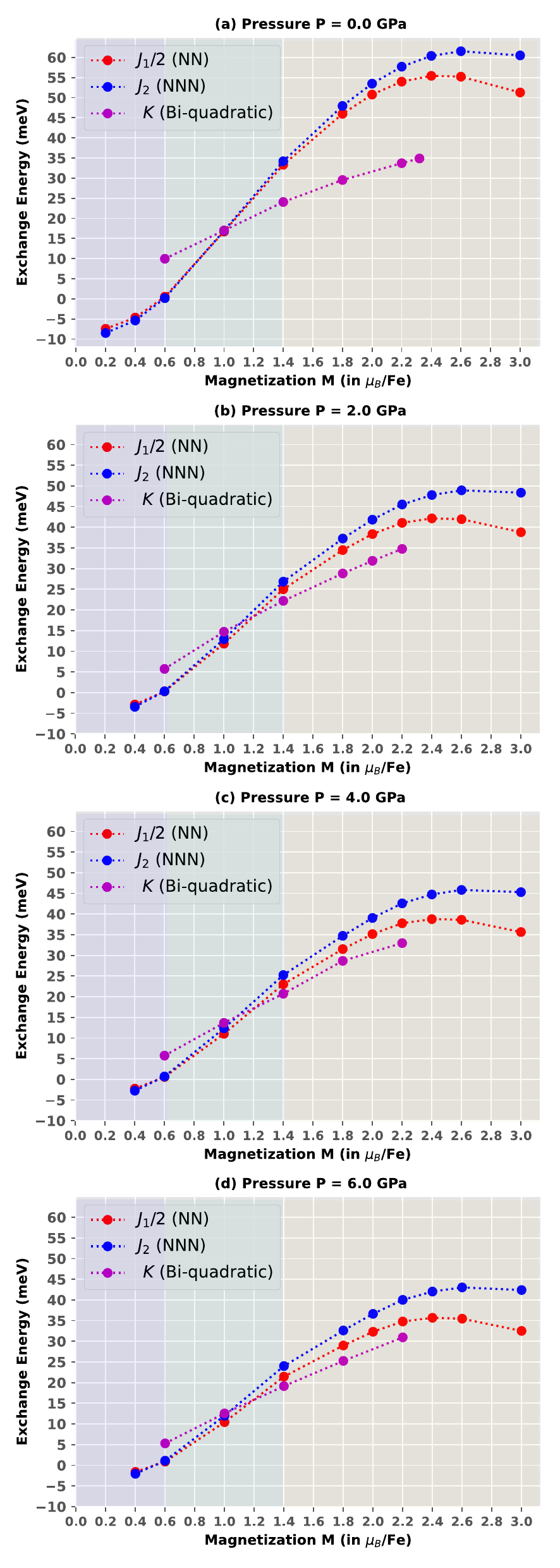}
 \caption{Dependence on magnetic moment ($M$) of the spin-exchange parameters $J_1$, $J_2$ and $K$ of the Heisenberg-like Hamiltonian in Eq.~\ref{1} calculated for different applied hydrostatic pressure $P$ (a) at ambient pressure ($P = 0$), (b) $P =$ 2 GPa, (c) $P =$ 4 GPa, (d) $P =$ 6 GPa. The values of $J_1$, $J_2$ are multiplied by $M^2$.} 
 \label{j}
\end{figure}

Figure~\ref{j} presents the dependence of the spin-exchange parameters on the magnetic moment at different applied hydrostatic pressures (we consider the {\em effective} value for the NN and NNN parameters, $J_i \times M^2, i=1,2$). Table~\ref{k1} summarises the results for two values of the magnetization: the optimal magnetization---defined as the magnetization corresponding to the minimum value of the magnetization formation energy for all the considered phases---and $M = 1.0\mu_B$---the magnetization close to which in Fig.~\ref{j}(a), $J_1/2 \sim J_2 \sim K$. 

For all pressures, $J_1/2$ and $J_2$, change from negative to positive when $M\gtrsim 0.6\mu_B$ (Fig.~\ref{j}). This corresponds to what is observed in Fig.~\ref{M}, where the FM phase is the most favourable in the LS state and AFM phases for the intermediate and HS state. For ambient pressure, this is consistent with an FM to AFM transition (in the SD configuration) as seen in Fig.~\ref{M}, while at higher pressure all phases are unstable in the LS state. $J_1/2$ and $J_2$ have the same behavior and take similar values for the observed range of magnetization, being nearly degenerate for intermediate magnetization values around $M= 1.0\mu_B$ (see also Table~\ref{k1}). This corresponds to the near degeneracy of the CAFM and SAFM phases (and SD at ambient pressure) in Fig.~\ref{M}. By increasing the pressure, the range of values that $J_1/2$, $J_2$ takes when varying the magnetization is reduced. The region where they are nearly degenerate is also reduced.  

At the optimal magnetization (HS state), $K$ (Table~\ref{k1}) is lower than $J_1(J_2)$, but still relatively large and supports different types of magnetic excitation like the SD phase. At $M = 1.0\mu_B$, $K$ is very close to $J_2$ and $J_1/2$. In fact, at ambient pressure and M=$1.0\mu_B$, the SD phase is energetically more favourable in comparison to other magnetic phases. The range of values taken by the biquadratic term $K$ shows relatively little changes with hydrostatic pressure (Fig.~\ref{j}). For all pressures, it increases from 5-10 meV for LS state to 30-35 meV in the HS state. A large $K$ according to the three-orbital Hubbard model~\cite{Mila2000,PhysRevLett.127.247204} originates from large hopping between unoccupied and occupied orbitals on neighboring magnetic ions. Its dependence on the ratio of the hopping parameters rather than their absolute values may explain the relatively little variation in $K$ as a function of pressure. As a consequence of the little variation with pressure, while at ambient pressure $K$ is remarkably different than $J_1/2$, $J_2$---being larger in the intermediate magnetization and almost half in the HS state---at higher pressures takes values in a similar range, as can be seen from Table~\ref{k1}. 
%------------------------------------------------------------
%----------------------------------------------------------
 The relative strength of the NNN and NN exchange couplings $J_2/J_1$ and of the biquadratic and NN exchange couplings $K/J_1$ can help interpret the phase diagram. $J_2/J_1$ is a measure for the competition between the CAFM and SAFM phases, $K/J_1$ indicates the presence of magnetic fluctuations in the HS state.
%------------------------------------------------------------------------------------------------------------------------------------
  \begin{table}[!htbp]
  \tiny
  \caption{Calculated exchange energies parameters ($J_1$, $J_2$, $K$ in Eq.~\ref{1}) at optimal magnetization (corresponding to HS state) and at $M = 1.0\mu_B$ (corresponding to LS state). Here, optimal magnetization is the magnetization corresponding to the minimum value of the magnetization formation energy for all the considered phases. The ratios $J_2/J_1$ and $K/J_1$ are also reported.}
  \label{k1}
  \resizebox{\columnwidth}{!}{
  \begin{tabular}{|lc|c|c|c|c|c|c|c|}
  \hline
%  \multicolumn{7}{|c|}{HS} \\ 
%  \hline
  &\text{P (GPa)}           & \text{M ($\mu_B$)}  & \text{$J_1$ (meV)} & \text{$J_2$ (meV)}  & \text{$K$ (meV)} & \text{$J_2/J_1$} & \text{$K/J_1$} \\ 
  \hline
 & \text{0.0}           &  2.3        & 110.82           & 60.37               & 34.66  &0.54   & 0.31 \\
 & \text{2.0}           & 2.0         & 76.70            & 41.83               & 31.88  &0.55   & 0.41 \\
 & \text{4.0}           & 1.9         & 70.30            & 39.04               & 28.67  &0.55   & 0.40 \\
 & \text{6.0}           & 1.8         & 57.94            & 32.61               & 25.25  &0.56   & 0.43  \\
   \hline
 %  \multicolumn{7}{|c|}{LS} \\ 
%    \hline
 & \text{0.0}      & 1.0         & 33.42             & 16.92                & 16.99 & 0.51     & 0.51 \\
 & \text{2.0}          & 1.0         & 23.66             & 12.91                & 14.71 & 0.55     & 0.62 \\
 & \text{4.0}          & 1.0         & 22.02             & 12.34                & 13.66 & 0.56    & 0.62 \\
 & \text{6.0}          & 1.0         & 20.84             & 11.98                & 12.54 & 0.57    & 0.60\\
 
  \hline  
  \end{tabular}
    }
  \end{table}
%------------------------------------------------------------------------------------------------------------------------------------

\begin{figure}
 \centering
 \includegraphics [width=0.45\textwidth]{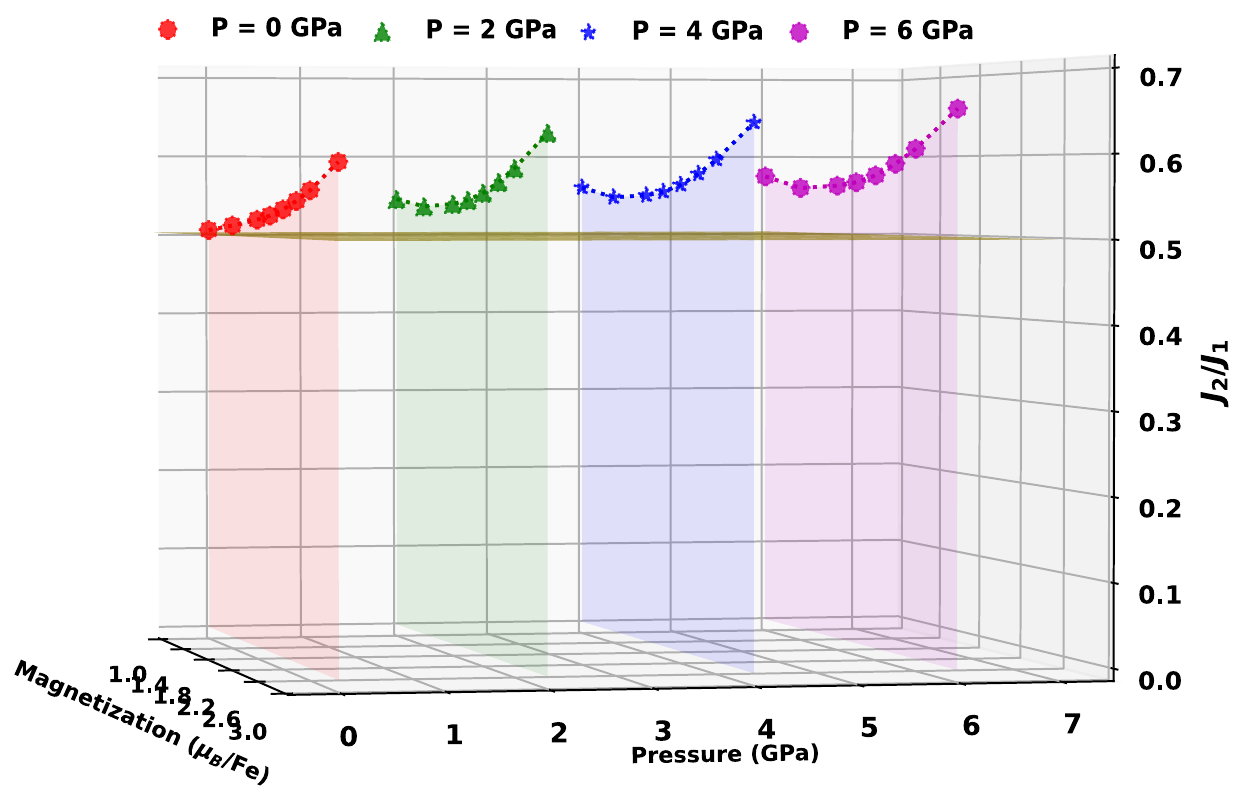}
 \caption{Dependence on magnetization $M$ of the ratio, $J_2/J_1$, between the NNN and NN spin-exchange parameters in Eq.~\ref{1} for different values of the applied hydrostatic pressure $P$. $P=0$ corresponds to ambient pressure.} 
 \label{k}
\end{figure}

Figure~\ref{k} presents the dependence of the relative strength of exchange coupling $J_2/J_1$ on pressure and magnetization for $M\geq 1.0\mu_B$. In the classical 2D mean field phase diagram the magnetic interactions $J_2/J_1 \approx 0.5$ corresponds to the boundary between the SAFM/CAFM phases.~\cite{PhysRevLett.101.057010} Within this framework, the results suggest that a SAFM/CAFM transition may be possible at ambient pressure and less likely at higher pressures. On the other hand, from a single-particle thermal occupancy viewpoint the region of lower magnetization is harder to reach at ambient pressure because the depth of the energy well associated with the AFM phase at ambient pressure is roughly twice that of the value observed at higher pressure (Fig.\ref{M}). Moreover, as the pressure increases the position of the minimum (in Table~\ref{mm}) is moving towards smaller magnetization.
%------------------------------------------------------------
%----------------------------------------------------------
 
\begin{figure}
 \centering
 \includegraphics [width=0.47\textwidth]{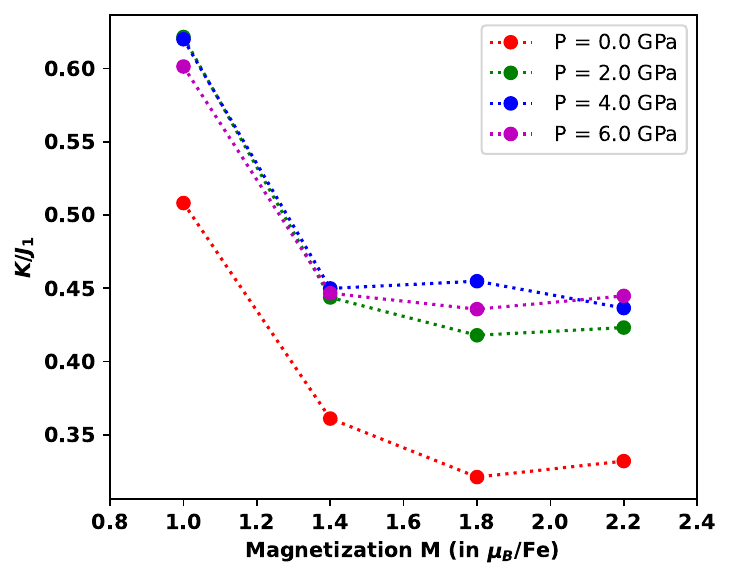}
 \caption{Dependence on magnetization $M$ of the ratio, $K/J_1$, between the biquadratic and NN spin-exchange parameters in Eq.~\ref{1} for different values of the applied hydrostatic pressure $P$. $P=0$ corresponds to ambient pressure.} 
 \label{m}
\end{figure}

Figure~\ref{m} presents the dependence of $K/J_1$ on magnetic moment and pressure is presented. At ambient pressure, a high value of $K/J_1$ ($\approx$0.31) in a high spin state means there is room for magnetic fluctuations. At higher pressure $K/J_1$ further increases. Considering the concurrent reduction of the energy difference between magnetic phases, this indicates that magnetic fluctuations are increasingly likely and strong. As a consequence, the mean-field theory approach is insufficient and one has to turn to methods that account for collective phenomena. In Sec.\ref{sec:QFTresult} we build on the DFT results of this and the previous section, using a field-theory approach to capture the effect of fluctuations. 
%----------------------------------------------------------
%----------------------------------------------------------

%\par One could also try to infer that $p=0$ case is closer to the QCP between two magnetic phases CAFM and SAFM. This is however misleading, because according to the Mermin-Wagner theorem due to fluctuations there is no long-range magnetic order in a 2D classical system with $J_1,J_2$ interactions. There was an effort, in \cite{PhysRevB.53.6455} to extend the 2D mean-field picture to 3D and to include quantum corrections in the large $S$ limit. It has been found that frustration extends also along the perpendicular direction, hence it is hard for a long-range order to settle. Even if we neglect this last finding and boldly compute $J_{\perp}$ from superexchange as a ratio $(t_{\perp}/t_{a,b})^2$ then we find a quantity that is order of magnitude smaller than $J_{1,2}$ found above. Then it is a problem of finding how far the QCP regime extends (and where the ordered and the classical critical regime are present). This problem has been solved recently in \cite{Frérot2019}, in particular Fig.4 therein, and analyzing their result leads us to conclusion that at temperatures $\sim 100K$ the system is classical (thermal) critical regime\footnote{And certainly it cannot be that at $p=0$ system is QCP while at finite $p$ system is ordered.}. This regime is addressed in Section\ref{sec:QFTresult} using field theoretical calculations.

\subsection{Orbital resolved Density of States}\label{ssec:HS-LS}
\begin{figure}[h!]
 \centering
 \includegraphics [width=8.5cm,height=11cm]{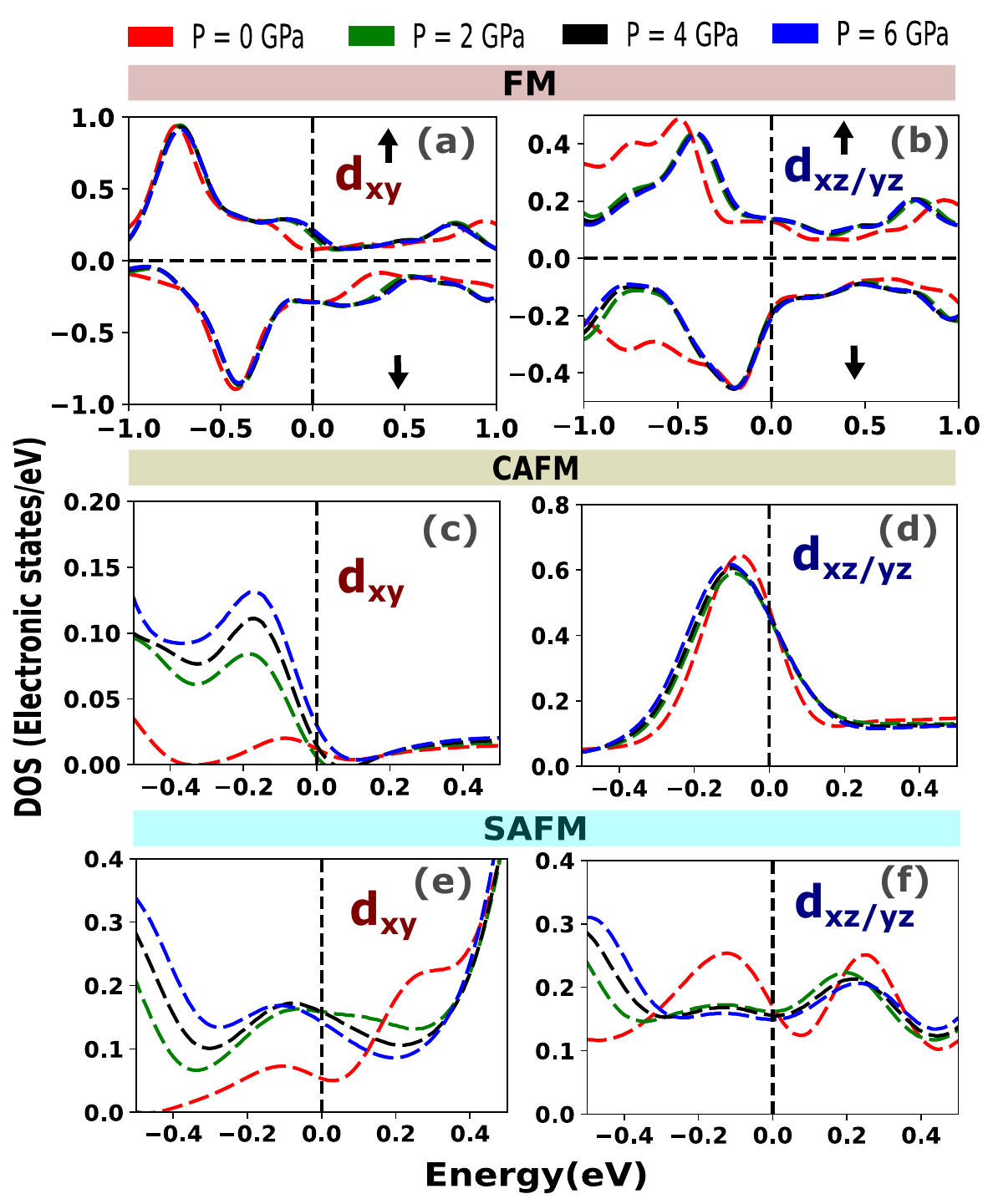}
 \caption{ Orbital resolved spin density of states (DOS) of $d_{xy}$ and $d_{xz/yz}$ in different magnetic phases - FM: (a), (b);  CAFM: (c), (d); SAFM: (e), (f) respectively, with pressure.}
 \label{dos}
\end{figure}

Figure~\ref{dos} presents the orbital resolved (spin-)DOS of the $t_{2g}$ $3d$ orbitals for the FM and CAFM and SAFM phases for different applied hydrostatic pressure. These orbitals are those which contribute the most at the Fermi surface (see Fig.~\ref{pdos}). 
In the FM phase, the orbital resolved spin-DOS is calculated at 0.2$\mu_B$ for which FM is predicted as the most stable phase (Fig.~\ref{M})---that is in a LS state. For the CAFM, SAFM phases,  the orbital resolved DOS is calculated at the phase optimal magnetization reported in Table~\ref{mm}---that is in an HS state.   
%------------------------------------------------------------
   %----------------------------------------------------------

 At ambient pressure, the $d_{xy}$ down-spin channel contributes dominantly to the occupied states near Fermi level of the FM phase [Fig.~\ref{dos}(a)].  As pressure is increased, there is an increment in partial DOS in up-spin channel. 
For the $d_{xz/yz}$ [Fig.~\ref{dos}(b)], the up/down channels contribution is similar at the Fermi level and there are no substantial changes when pressure is increased. Looking at the overall energy range, there is a shift in the peak for the up-spin channel when increasing pressure that corresponds to a reduction in the spin-exchange parameters (see Fig.~\ref{j}). This indicates reduced FM fluctuations with increased pressure.   

Figure~\ref{dos}(d) shows that there is a significant contribution from only $d_{xz/yz}$ orbital in the partial DOS at Fermi level in CAFM phase at ambient pressure. The partial DOS of $d_{xy}$ orbital at Fermi level is negligible [Fig.\ref{dos}(c)].  Increasing the pressure does not change the contribution at the Fermi level substantially. With pressure, the peak below the Fermi level of $d_{xz/yz}$ is slightly reduced and shifted to lower energies and  the smaller peak below the Fermi level for  $d_{xy}$ increases. 

At ambient pressure, in the SAFM phase, both the $t_{2g}$  and $e_{g}$ orbitals contribute to the DOS at the Fermi level [Fig.\ref{pdos}(C)]. The $d_{xz/yz}$ has the highest contribution whereas the $d_{xy}$ has the lowest. Increasing pressure, the contribution $d_{xy}$ becomes roughly equal to that of  $d_{xz/yz}$ while the contribution from $e_g$ orbitals is suppressed [Fig.~\ref{pdos}(C)]. 

The overall picture that emerges from the evolution of partial DOS with pressure is the exclusivity of the $d_{xz/yz}$ orbitals in the CAFM phase and conversely, the key role played by the $d_{xy}$ orbital in the FM and SAFM. In particular, the enhanced $d_{xy}$ partial DOS with pressure correlates with the reduction of FM fluctuations in the LS phase and the emergence of long-range SAFM order in the HS state. 

\section{2D effective field-theory model}\label{sec:QFTresult}

Experimentally, long-range spin order is observed between 2-6 GPa. While the DFT calculations correctly predict the disappearance of long magnetic order around 6 GPa, they showed a propensity towards the formation of sAFM order below 2 GPa in contrast with the experimental findings.  In particular, DFT results indicate that i) the bandwidth of the $d$ orbitals at the Fermi energy is the closest to the experiment when long-range magnetic order is considered and ii) the value of spin-exchange parameters $J_{1,2}$ is larger while $K/J$ is smaller at zero than at finite applied pressure. In what follows, we put forward a classical two-dimensional (2D) spin model to reconcile the DFT results with the experimental findings.   

Using a 2D model is justified by FeSe being a layered material, with a strong anisotropy of the spin-exchange parameters $J$, which is reflected in the spin anisotropy. The classical treatment of the spin degrees of freedom is justified because, due to its multi-orbital nature, FeSe is neither fully itinerant nor fully spin-localized. Then, a fraction of spin is localized, but spins can gradually change their amplitude and orientation due to free flow back and forth into the itinerant bath, which can be described classically~\cite{Temme2012}.

\subsection{Hamiltonian for the orbital degrees of freedom}

The DFT calculations show a strong dependence of the spin-exchange parameters on the local on-site magnetization, which points to a strong coupling between the spin and orbital degrees of freedom: varying the on-site magnetization is equivalent to selecting the spin configurations (high-spin or low-spin) of the Fe $d$-orbitals. Then, to model the high-spin or low-spin dependence into the effective low-energy theory, we introduce a Hamiltonian for the $d$ orbital degrees of freedom. We take fermionic annihilation operators $c_{\sigma\xi}(i)$ at a site $i$ with spin $\sigma$ and orbital $\xi$ indexes and notice that the fluctuation of orbital content by analogy with spin space operators will correspond to a bosonic operator defined as $b_i=c^{\dag}_{\sigma\xi}(i)c_{-\sigma\xi\pm 1}(i)$.
From Fig.~\ref{M}, we observe that the energy difference between the lowest magnetic and the non-magnetic configuration can be fitted as a cosine of the spin-magnetization $|S|$, $- W_L \cos(\delta|S|)$ with $\delta|S|=|S|-|S|_0$. Remarkably, $W_L$ strongly depends on pressure, with $W_L$ at ambient pressure being approximately twice the $W_L$ at 2 GPa.

We then consider a basis of on-site localised bosonic states associated with the $d$-orbital fluctuations and define the following Hamiltonian in terms of creation and annihilation operators $b_{i}, b_{i}^\dag$  
\begin{equation}
    H_\text{orb} = -W_L \sum_{i} (b_{i}^\dag b_{i+1}^{}+h.c.) + J_H \sum_i b_{i}^\dag b_i^{} b_{i}^\dag b_i^{},
\end{equation}
with a next-neighbor hopping-like term (strictly speaking the cosine dependence $- W_L \cos(\delta|S|)$ defines an analog of a $\xi-$distortion potential and then following analogy with elasticity we define "elastic" modes of fluctuation in orbital space, we take that the origin of these is locally modified tight-binding parameter $\delta t \sim \delta |S|$) $W_L$ and a quadratic $J_H$ Hund's exchange term.\cite{sym13020169} In this simple tight-biding model, the hopping parameter is exactly the high-spin/low-spin energy difference that can be fitted from the curves in Fig.~\ref{M}. The cosine fit works well, so we can restrict ourselves to nearest-neighbor tight-binding approximation, although in principle field theory does not require it. 

Since orbital fluctuations are confined to the 2D plane and can be assumed as continuous, classical variables, we follow Villain\cite{Villain-trans} to re-write the Hamiltonian in terms of bosonic fields:
\begin{equation}\label{eq:orbit-bosoniz}
%   H_{orb} = \int dr W_L (\nabla\theta_L)^2 + J_H (\nabla\phi_L)^2 + g_L\cos\phi_L
   H_\text{orb} = \iint dx\,dy \left(W_L (\nabla\theta_L)^2 + J_H (\nabla\phi_L)^2 + y_L\cos\phi_L\right)
\end{equation}
where $\theta_L(x,y)$ is the orbital field, related to the local density of $b$-bosons, $\nabla_x\theta_L\propto b^\dag(x)b(x)$, $\phi_L(x,y)$ is the canonically conjugated field and $y_L$ is the fugacity. The fugacity parameter, $y_L \propto \exp(-\epsilon\beta W_L)$, is related to the temperature $1/\beta$ and the low-spin/high-spin energy difference $W_L$.\cite{Giachetti2023} The factor $\epsilon$ is the analog of the relative permittivity in 2D electric gas. 
%and the Hund's exchange $J_H$ that plays a role of interactions . 
The key difference with the usual Villain Hamiltonian is that due to the presence of strong correlations, the fluctuations of canonically conjugated $\phi_L$ field appear, which modifies the value of $\epsilon$.\cite{Kovchegov_2003} Namely, in the original Kosterlitz and Thouless picture $\epsilon=\pi^2/2$, including vortex screening\cite{Benfatto_12} gives $\epsilon=3/\pi$, while including the canonically conjugated term, one that is proportional to $J_H$, gives an extra factor $(W_L/J_H+2)/(2(W_L/J_H+1)$ to $\epsilon$. The cosine term in Eq.~\ref{eq:orbit-bosoniz} has been introduced in the 2D context by Villain\cite{Villain-trans} to capture low-energy large-angle fluctuations. Because of the latter term, the Hamiltonian hosts vortex excitations. The vortex excitations manifest physically as regions of gradually lower spin magnetization in an overall high-spin background.
%In the language of Coulomb gas, vortex excitations correspond to magnetic fluctuations responsible for the renormalization of the effective temperature.~\cite{Giachetti2023}\footnote{the theory is a relativistic theory in a moving system of coordinates which causes dilatation of all characteristic lengths, this is an underlying reason for the ``temperature'' renormalization}

\begin{figure}[h]
 \centering
 \includegraphics [width=0.45\textwidth]{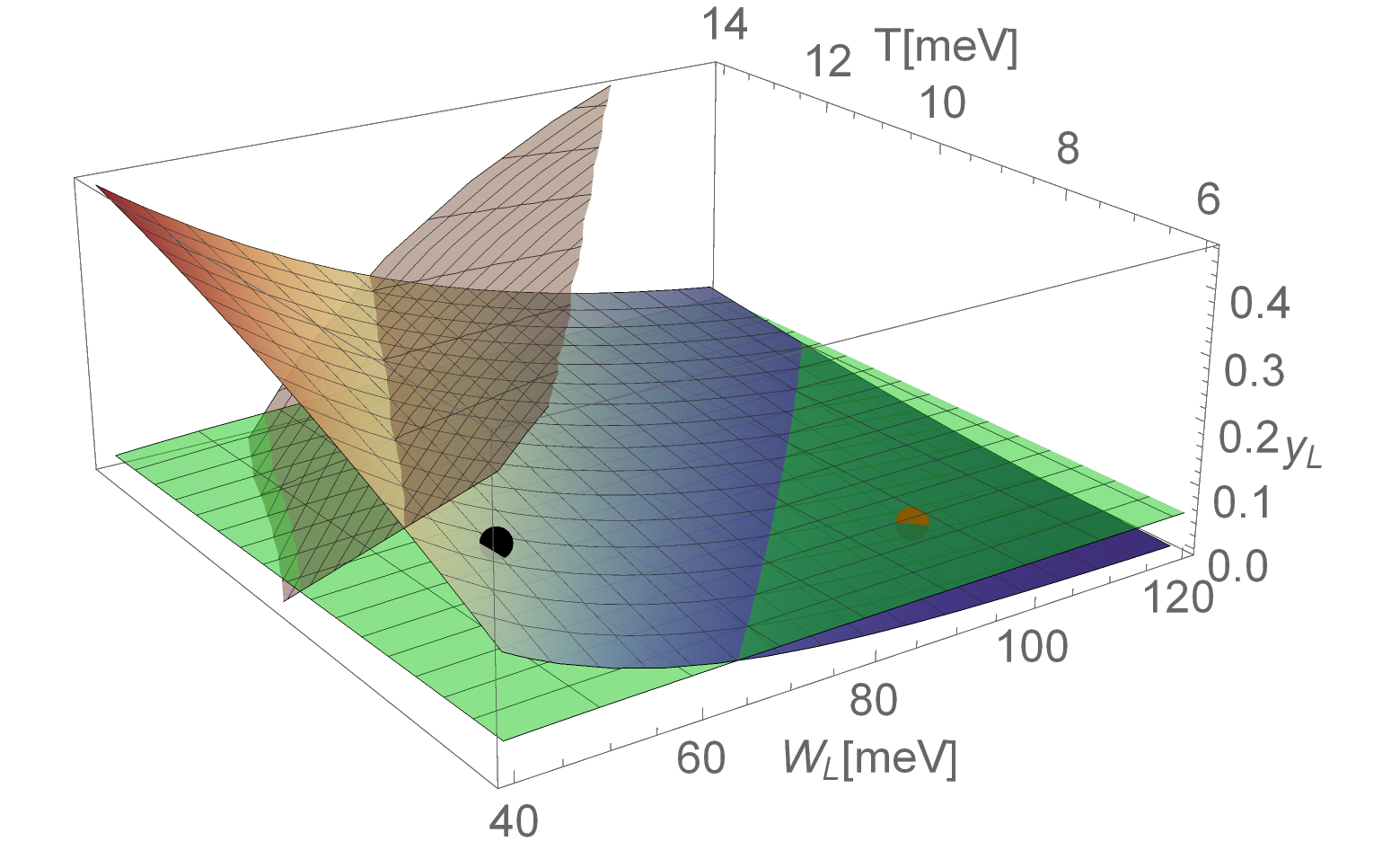}
 \caption{Shaded blue-to-red surface show, for our chosen $J_H$, the orbital vortex fugacity $y_L$ as a function of orbital fluctuations bandwidth $W_L$ and temperature $T$. Green and gray planes are the lower and upper limits of the vortex crystal phase, respectively. We show the location of $p=0$ (red dot) and $p\geq 2$GPa (black dot). The red dot is just below the green plane hence in the regime where bound pairs of vortexes exist but become dilute in the thermodynamic limit ($x,y\rightarrow\infty$), while the black dot is well inside the vortex crystal phase where topological orbital excitations are ordered and their number is constant.}
 \label{fig:fugacityL}
\end{figure}

The value of $y_L$ determines the system's behavior.\cite{PhysRevLett.71.2138,RevModPhys.59.1001} This is illustrated in Fig.~\ref{fig:fugacityL} for $J_H =$ 0.6 eV---which was calculated for FeSe---\cite{sym13020169} where $y_L$ is shown as a function of the temperature $1/\beta$ and $W_L$. In the region below the green plane ($y_L = 0.054$) the system is below the vortex regime and physics is dominated by density waves. Above $y_L = 0.054$, the vortex system undergoes a crystallization transition. Long-range magnetic order is then allowed in the high-spin regions between vortexes.
In the region between the green and the gray planes, the vortexes form a rigid network, long-range magnetic order is allowed in the high-spin regions between vortexes. Above the gray plane, vortexes move freely and static long-range magnetic order is destroyed. Extracting the values of $W_L$ from Fig.~\ref{M}, we found that slightly above 100K, FeSe is in the density wave region at $p=0$, while at $p=2$ GPa is in the vortex crystal phase. At larger pressure and lower $W_L$, the system is in the free vortex regime. 

This result from the model agrees with the experimental observation of long-range magnetic order being observed in FeSe for intermediate pressures, while absent at ambient pressure and above 6 GPa. Further, the above model provides a rationale for the tendency in the DFT results to long-range magnetic order at ambient pressure which is not observed experimentally. The long-range magnetic order is suppressed by the orbital vortex formation because of the coupling of the spin with "randomness" present in the orbital degrees of freedom. In Fig.\ref{fig:fugacityL} we show that this randomness is present at ambient pressure, but at higher pressure the orbital sector orders.
The suppression of the long-range spin order by vortex formation cannot be captured in standard DFT calculations on a (magnetic) unit cell and the long-range magnetic order is favored over the non-magnetic configuration both energetically and when considering the electronic structure close to the Fermi level. On the other hand, when DFT on large supercells are used, it has been shown~\cite{PhysRevB.102.235121} that random-spin configurations are energetically more favorable than the non-magnetic ordered ones, in agreement with the model above.    

\subsection{Hamiltonian for the spin degrees of freedom}

As noticed previously, the strong dependence on the spin-magnetization of the spin-exchange coefficients of the reduced Heisenberg Hamiltonian in Eq.~\ref{1} indicates a strong coupling of the spin and orbital degrees of freedom. Considering the above model for the orbital degrees of freedom (Eq.~\ref{eq:orbit-bosoniz}), the local variations of $\nabla\phi_{L}$ modify the local parameters of spin fluctuations $\langle\nabla\phi_L(x_i)\rangle\neq 0 \implies \delta J(x_i)$ and hence the local energy of the spin system.  In a mean-field picture $J(x)S(x)S(x\pm1)\approx J_{avr}S(x)S(x\pm1)+\delta J(x)S(x)\langle S(x)\rangle$, namely the spin-exchange parameters $J$ result by averaging over the orbital degrees of freedom and the variations of $\nabla\phi_{L}$ result in an effective magnetic field $h(x,y)=\delta J(x)\langle S(x)\rangle$. Then, the Hamiltonian for the spin degrees of freedom coupled with the orbital degrees of freedom consists of a Heisenberg-like model, including a quadratic term (similar to Eq.~\ref{1}), and an additional term depending on $h(x,y)$:   

\begin{align} \label{eq:SLmod} 
    H_{s(+L)}&= \sum_{ij}J_{ij}\vec{S}_i\cdot\vec{S}_j - K\sum_{i}(\vec{S}_i\cdot\vec{S}_{i\pm 1})^2+\\ 
    &+ h(x,y)(S^x(x,y) + S^y(x,y)). \notag
\end{align}
The spatial distribution of $h(x,y)$ in Eq.~\ref{eq:SLmod} may be either random (for orbital density fluctuation regime at $P=0$) or periodic (for vortex crystal, $P\geq 2$GPa). In the latter case, one expects the opening of Bragg-mini gaps in the spin excitation spectrum at appropriate magnetic reduced Brillouin zone boundaries. In the former case, one expects rare, randomly positioned areas of the low-spin state which will induce a variation of $J(n_L)$ hence disorder-induced localization. This is true not only when the potential associated with $h$ changes abruptly and thus the long-range spin-order is destroyed by backscattering, but also for a sufficiently smooth potential, such that forward scattering dominates. It can be shown\cite{Giam88} that $h(x,y)$ introduces exponential decay pre-factor in front of any spin-spin correlation functions, and thus only a short-range spin-order is possible. A more advanced and quantitative treatment of spin and orbital fluctuations based on the full solution of the renormalization group equation is postponed to further study.         
%Obviously, this is the simplest mean-field picture, the precise collaboration between spin and orbital degree of freedom at each distance would require more advanced methods, for instance deriving renormalization group (RG) equations for the problem. The proof, using RG methods, that not only random vortex suppresses spin-order but also that orbital order can indeed enhance order in the spin-sector, will be postponed from here to the follow-up paper. ADD THIS IN THE CONCLUSIONS 

\subsection{Renormalization of the electronic structure}

The orbital fluctuations from Eq.~\ref{eq:orbit-bosoniz} couple with the electronic degrees of freedom and can renormalize the electronic structure (e.g. from DFT calculations) by reducing the band dispersion. The coupling between a propagating electron and a boson describing local fluctuation of orbital content can be treated like any electron-boson coupling. In particular, electron-phonon coupling leading to the emergence of polarons is the most explored, and we shall use those results here to obtain band renormalization. We follow the standard Feynmann path integral prescription to compute the polaronic effect, but with the boson-boson interaction accounted for. Since we do not know the amplitude of electron-boson coupling, $g_{el-L}$ we take it as a free parameter and plot the reduction of the electron velocity $\alpha_V$ (thus a reduction of the band curvature) as a function of $g_{el-L}$. In principle, this coupling can be determined from first principles following Ref.~\cite{PhysRevB.94.115208}.

\begin{figure}
 \centering
 \includegraphics [width=0.45\textwidth]{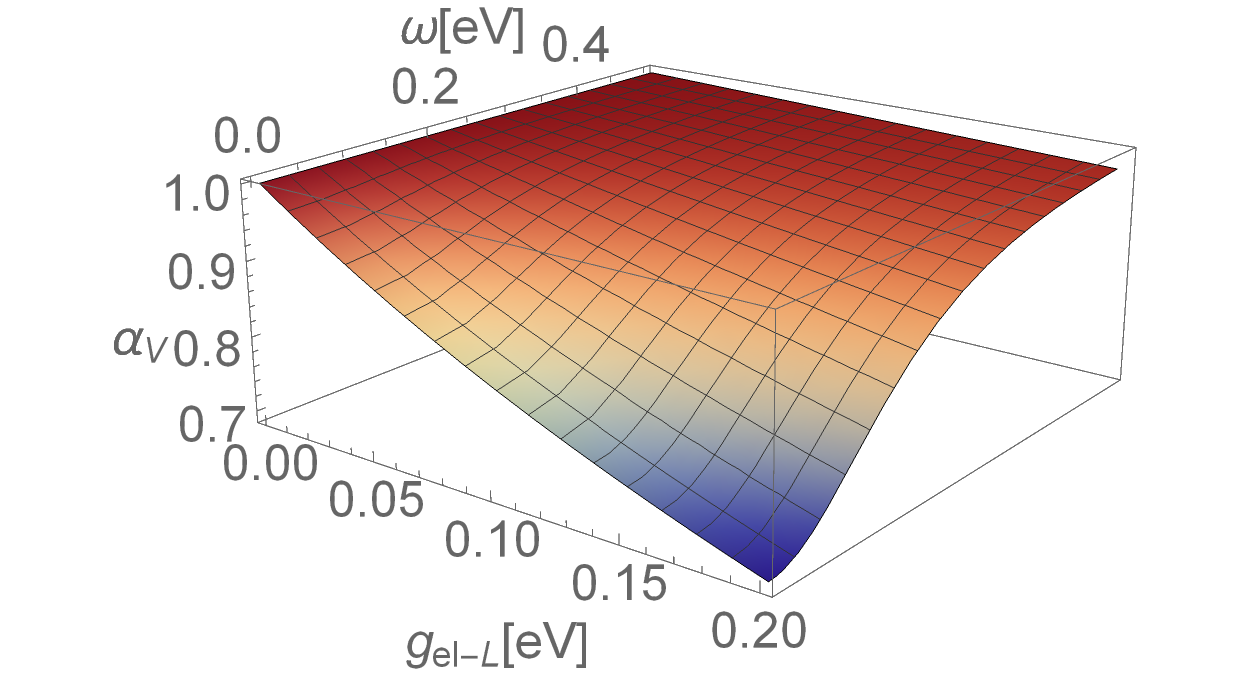}
 \caption{Renormalization factor $\alpha$ of velocity as a function of coupling between single carriers and orbital fluctuations $g_{el-L}$ and distance from chemical potential $\omega$.} 
 \label{fig:alpharen}
\end{figure}

The result is presented in Fig.~\ref{fig:alpharen}. Close to Fermi energy ($\omega =0$ eV), the observed velocity can be reduced by $30$\% for $g_{el-L} \sim 0.2$. As one moves away from the Fermi energy the renormalization factor goes to 1---no renormalization. The strong energy dependence implies that the effect should be distinguishable from the Hubbard $U$ mechanisms of bandwidth renormalization and that care needs to be taken if one attempts to infer the bandwidth from a partial dispersion relation.    

We expect the renormalization to be larger for the band with $d_{xy}$ character because of concurrent effects. First, this band has the smallest DFT dispersion and the dispersion enters into the denominator of the dimensionless coupling $\propto g_{el-L}$. Second, by construction, the orbital field $\theta_L$ in Eq.~\ref{eq:orbit-bosoniz} is zero at high spin and maximum at low spin state when the occupancy of the  $d_{xy}$ orbital changes from $1$ to $0$ (while the occupancy of the degenerate $d_{xz,yz}$ changes by a fraction between $1$ to $2$). Then, $d_{xy}$ is most affected by orbital fluctuation and as such we expect a larger reduction in the bandwidth than for $d_{xz,yz}$. This is consistent with what was observed in ARPES experiments. 

For the above-given renormalization effect, we only considered the first two terms in Eq.~\ref{eq:orbit-bosoniz}, i.e. orbital occupancy density waves. 
In principle also electrons couple to the vortexes described by the last term in Eq.~\ref{eq:orbit-bosoniz}. The reason we did not consider this last term is that a theory for such coupling is not available, although one can anticipate that the coupling is the largest for electron dispersion close to the $\Gamma$ point (due to the $k-$dependence of the Fourier-transform of a solitonic wave) and that there are several satellites separated by $\sim 0.1$~e, which may be visible in experimental spectra.

\section{Conclusions}
Using DFT at the generalized gradient approximation level, we calculated the dependence of the magnetic formation energy of FeSe on the local magnetization for different magnetic phases and we mapped the results into a Heisenberg-like Hamiltonian with two main outcomes. First, we observed a strong dependence of the spin-exchange parameters extracted from the Heisenberg-like Hamiltonian on the local on-site magnetization which points to a \emph{strong coupling between spin and orbital degrees of freedom}. Second, we obtained stable antiferromagnetic orderings at ambient pressure. The latter result is consistent with the results we obtained for the electronic structure, where assuming an antiferromagnetic ordering renormalizes the bandwidth of the $d$-bands, bringing it closer to the ARPES results. On the other hand, this propensity towards an antiferromagnetic phase is in disagreement with experimental results where magnetic ordering is observed only for a pressure of 2 GPa or larger.

We argued that though a thermodynamically stable long-range magnetic order at ambient pressure has not been detected, the DFT results may indicate the existence of a short magnetic order convoluted with slow variations of spin amplitude at long range. Such an inhomogeneous spin pattern would emerge from the strong coupling between the spin and orbital fluctuations and from the quasi-2D nature of FeSe.
Indeed, we showed that the DFT results can be reconciled with the experimental observations within a \emph{2D effective field theory which admits non-trivial, spatially extended topological vortex states}. According to the model, the formation of vortexes both suppresses the antiferromagnetic phase at ambient pressure and plays a role in renormalizing the bandwidth of $d$-band. The existence of mesoscopic structure such as vortexes is comforted by the recent observation of Griffiths phases~\cite{PhysRevLett.127.246402} in FeSe$_{1-x}$S$_x$.

This 2D effective field-theory model implies the impossibility of capturing the bandwidth renormalization of FeSe close to the Fermi energy uniquely by improving the description of electron correlation beyond DFT, %HERE WOULD BE NICE TO BE MORE SPECIFIC ON CORRELATION TYPE - THIS IS VAGUE
and instead points to the need of accounting for spin and orbital fluctuations at the mesoscopic scale. Our DFT results for the electronic structure hint that to partly reproduce such effect---without resorting to large supercells as in Ref.~\cite{PhysRevB.102.235121}---a ``poor man"-approach would be to assume an antiferromagnetic ordering. A more sound approach is to account for the interaction between electrons and the fluctuations via a polaronic-like model. Our preliminary results show that the electronic structure can be indeed renormalized up to 30\% and the most affected band would be that with $d_{xy}$ character.

Finally, the 2D effective field-theory model presented here---or rather a refined version of it based on the full solution of the renormalization group equation---together with the results on the existence of an SD magnetic ordering---which we found to be energetically favored for intermediate magnetization---can be used to investigate the origin of the nematic phase in FeSe. %PROVIDE SOE CONTEXT DEBATE ORBITAL VS SPIN OR BOTH

\section{Acknowledgements}
This work was supported by the Engineering and Physical Sciences Research Council (EPSRC), under grant EP/V029908/1. The authors are grateful for the use of the computing resources from the Northern Ireland High Performance Computing (NI-HPC) service funded by EPSRC (EP/T022175); and from the UK Materials and Molecular Modelling Hub partially funded by EPSRC (EP/T022213/1, EP/W032260/1 and EP/P020194/1)). The authors acknowledge useful discussions with Nigel Hussey, Antony Carrington, Swagata Acharya and Marc Gabay.
%{\bf |MORE NAMES?}

\input{appendixes}

\bibliography{FeSe}
\end{document}

%% file: appendixes.tex
\appendix

\section{Evolution of structural parameters with pressure in different magnetic phases}

%------------------------------------------------------------
\begin{figure}
 \centering
 \includegraphics [width=0.45\textwidth]{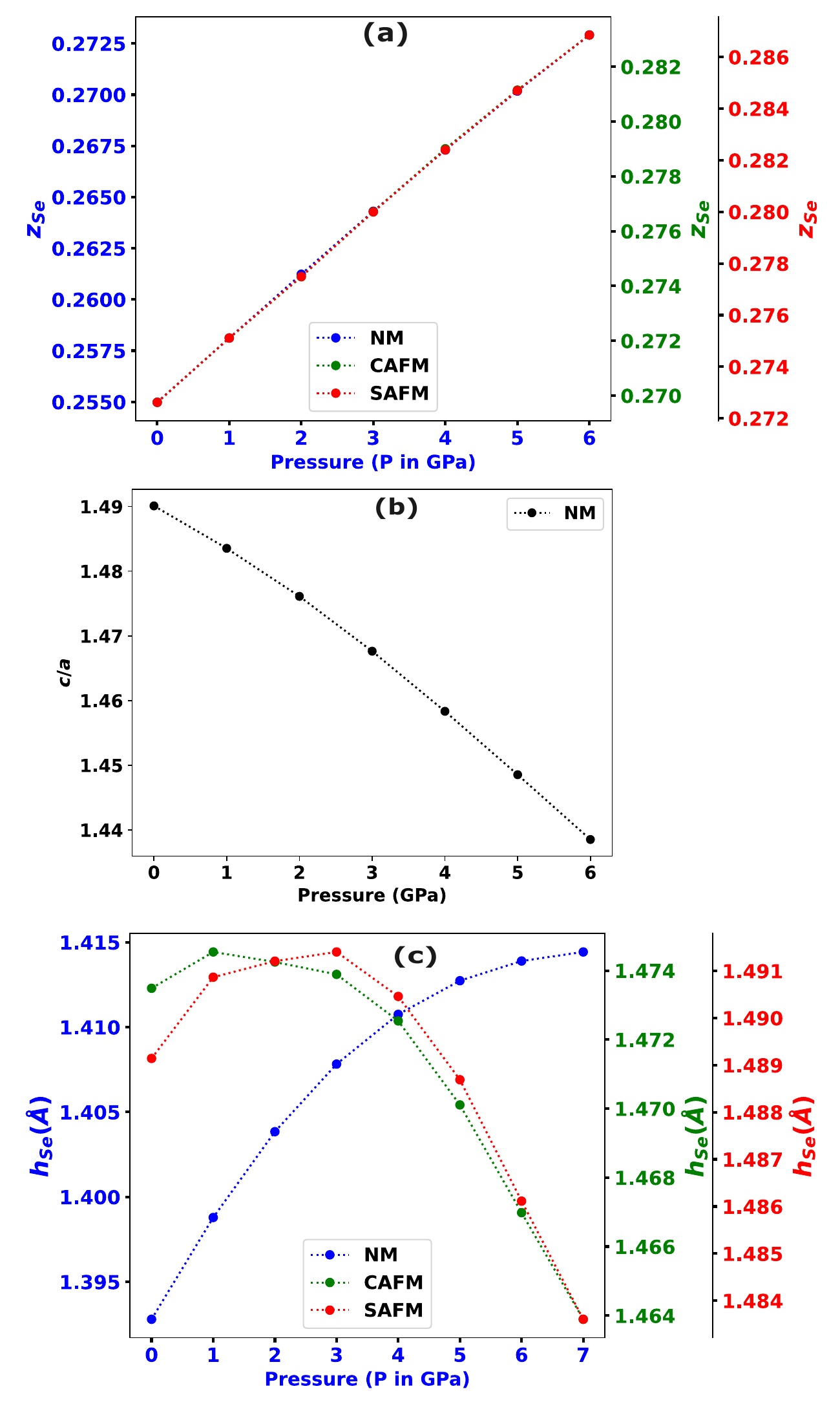}
 \caption{ Variation of (a) $z_{Se}$, (b) $c/a$ and (c) $h_{Se}$  with pressure in non-magnetic (NM), checkerboard/Neel antiferromagnetic (CAFM), and striped AFM (SAFM) phase.}
 \label{zse}
\end{figure}
   %----------------------------------------------------------
  
Evolution of different structural parameters like internal $z$ position of Se ($z_{Se}$), $c/a$ ratio and height of the Se-atom from the Fe-plane ($h_{Se}$) are presented in Fig.\ref{zse}. The comparison of our DFT calculated structural parameters with experiment is presented in Table \ref{s}. A linear increase in $z_{Se}$ with pressure in non-magnetic phase and with different long range magnetic order is very much evident from Fig.\ref{zse}(a). At ambient pressure, the calculated value in non-magnetic phase is very close (slightly underestimated) to experimental as well as  DFT+DMFT calculated value \cite{PhysRevB.97.115165,doi:10.1021/jp1060446,B813076K}. Introducing long range order into the system slightly overestimates the value. At higher pressures (4GPa), the experimental $z_{Se}$ is very close to the calculated value with SAFM long range order. Looking at the lattice parameters reveal a decreasing trend in $c/a$-ratio with pressure (Fig.\ref{zse}(b)) which is consistent with experiments. Calculated $c/a$ at ambient pressure is overestimated by 2\% with respect to the experimental value and very close to DFT+DMFT results. The lattice parameters are kept fixed for the calculations with long range magnetic order. Anion height ($h_{Se}$) has been found to be an important factor controlling magnetism as well as superconductivity in iron based superconductors \cite{Mizuguchi_2010}. Therefore, it is worth investigating the variation of $h_{Se}$ at different magnetic phases with external hydrostatic pressure. Fig.\ref{zse}(c) shows the variation of $h_{Se}$ with pressure. The variation in NM, CAFM and SAFM phases are presented with blue, green and red colors respectively. If we look at the NM phase, it is conspicuous that there is a gradual increase in $h_{Se}$ with pressure. A steep increase is evident at lower pressure values which ultimately reaches almost saturation at pressure greater than 5.0 GPa. Magnetic interactions seem to increase $h_{Se}$ to a great extent. The calculated value of $h_{Se}$ at ambient pressure with optimized structure is 1.3927 $\AA$ in non magnetic phase. The value is underestimated by nearly 4\% in comparison to experiments. $h_{Se}$ is increased to 1.4734 $\AA$ and 1.4891 $\AA$ in CAFM and SAFM phase respectively. The situation improves as long range order is introduced into the system and the value is within 2\% of overestimation. The variation in CAFM phase is marked by an increase in $h_{Se}$ at P = 1.0 GPa, beyond which it is decreased gradually. The rate of decrement is faster at pressure greater than 4.0 GPa. In case of SAFM, the nature of variation is somewhat different in comparison to CAFM. A sudden increase in $h_{Se}$ is followed by a plateau upto pressure 3.0 GPa. A gradual reduction in $h_{Se}$ just like CAFM is observed beyond 3.0 GPa. From the above discussion, it is evident that the structural parameters calculated via our DFT approach exhibit strong agreement with experimental results, both in non-magnetic and long-range magnetic ordered FeSe.

%------------------------------------------------------------------------------------------------------------------------------------
  \begin{table}
  \tiny
  \caption{Comparison of structural parameters}
  \label{s}
  \resizebox{\columnwidth}{!}{
  \begin{tabular}{|lc|c|c|c|c|c|}
  \hline
  &\text{Structural parameters}           & \text{DFT(NM)}     & \text{DFT+CAFM}     & \text{DFT+SAFM}       & \text{Experiment}       \\ 
  \hline
  & \text{$z_{Se}$ (0 GPa)}           & 0.2550             & 0.2697              & 0.2726                & 0.2660              \\
  & \text{$z_{Se}$ (4 GPa)}           & 0.2673             & 0.2790              & 0.2823                & 0.2915            \\
   \hline
  & \text{$c/a$    (0 GPa)}           & 1.4900             & -                   & -                     & 1.4580            \\
  & \text{$c/a$    (4 GPa)}           & 1.4583             & -                   & -                     & 1.4215            \\
   \hline
  & \text{$h_{Se}$ ($\AA$)    (0 GPa)}           & 1.3927               & 1.4734              & 1.4891                & 1.4502              \\
  & \text{$h_{Se}$ ($\AA$)   (4 GPa)}           & 1.4107               & 1.4725              & 1.4904                & 1.4233              \\
  \hline  
  \end{tabular}
    }
  \end{table}

%------------------------------------------------------------------------------------------------------------------------------------

\section{Cosine fits $J_{1,2}$}

One very interesting observation is that \textit{effective} $J_1$, $J_2$ curves can be well-fitted with a cosine(M) function like $w[-\cos({aM+\alpha})+c]$, where $w$ is \enquote{width} and $c$ is \enquote{offset} of the fitted curve. The absolute value of the parameter $w$ decreases with increase in pressure. This denotes decreased $J_{eff}$ with increased pressure. On the other hand, the parameter $c$ favors AFM over FM. A gradual increase in absolute value of $c$ is observed with increased pressure. This indicates possible FM fluctuation at ambient pressure and higher stability of AFM state at higher pressure in FeSe. 
%------------------------------------------------------------
\begin{figure}
 \centering
 \includegraphics [width=0.35\textwidth]{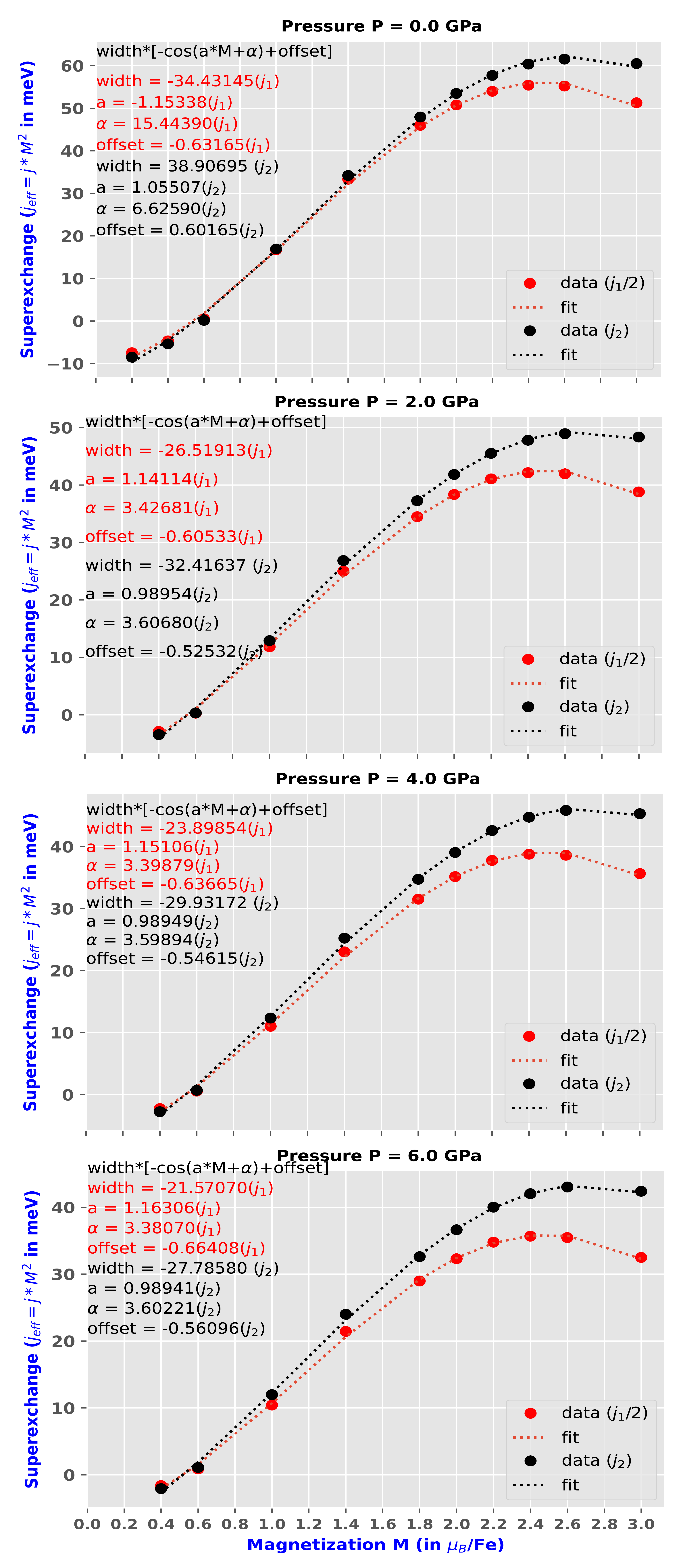}
 \caption{ Cosine fitting of $J_1$ and $J_2$ curves.}
 \label{cosfit}
\end{figure}
%----------------------------------------------------------

\section{Bi-quadratic $K$ from non collinear calculation}
%------------------------------------------------------------
\begin{figure}
 \centering
 \includegraphics [width=0.4\textwidth]{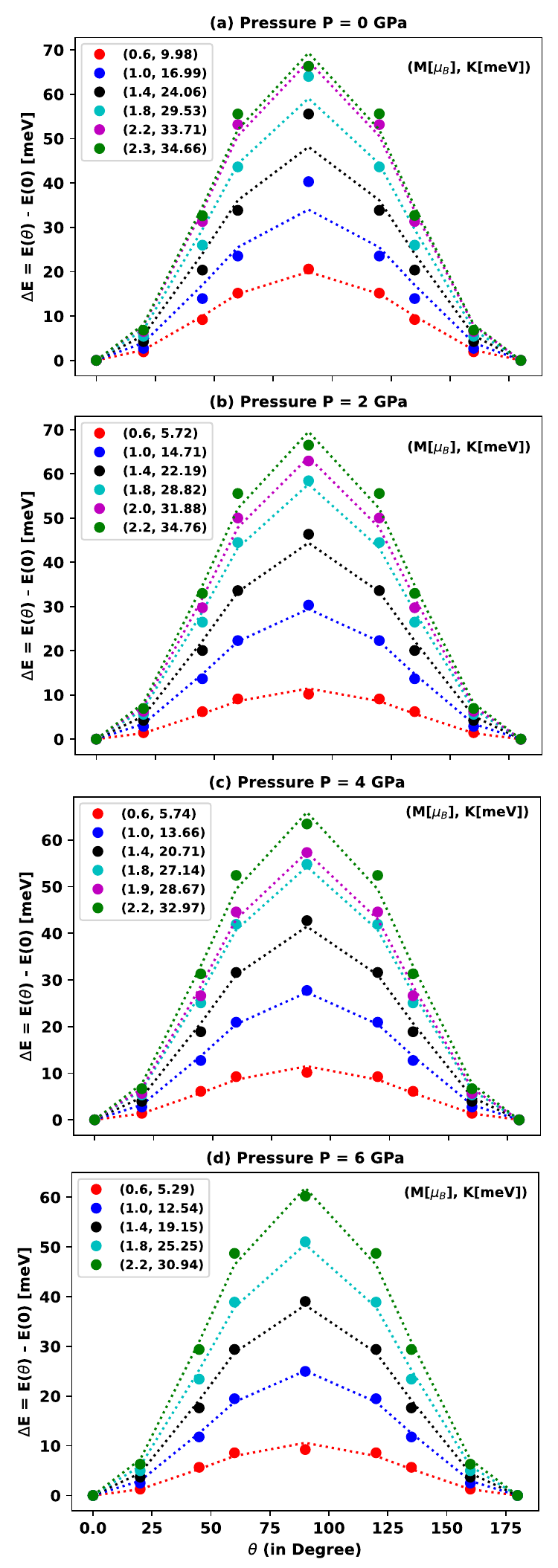}
 \caption{ Bi-quadratic $K$ at different pressure and magnetic moment.}
 \label{fit}
\end{figure}
%----------------------------------------------------------
The values of $\Delta E(\theta)$ calculated from non-collinear calculations and their fitting with equation.\ref{4a} at different magnetic moments are presented in Fig.\ref{fit}. Simple least square fitting method is incorporated to extract the value of $K$ at each magnetic moment. It is evident from Fig.\ref{fit} that we get a better fit at high spin states in comparison to the low spin ones.

\section{Spin resolved partial DOS}
%------------------------------------------------------------
\begin{figure*}
 \centering
 \includegraphics [width=0.90\textwidth]{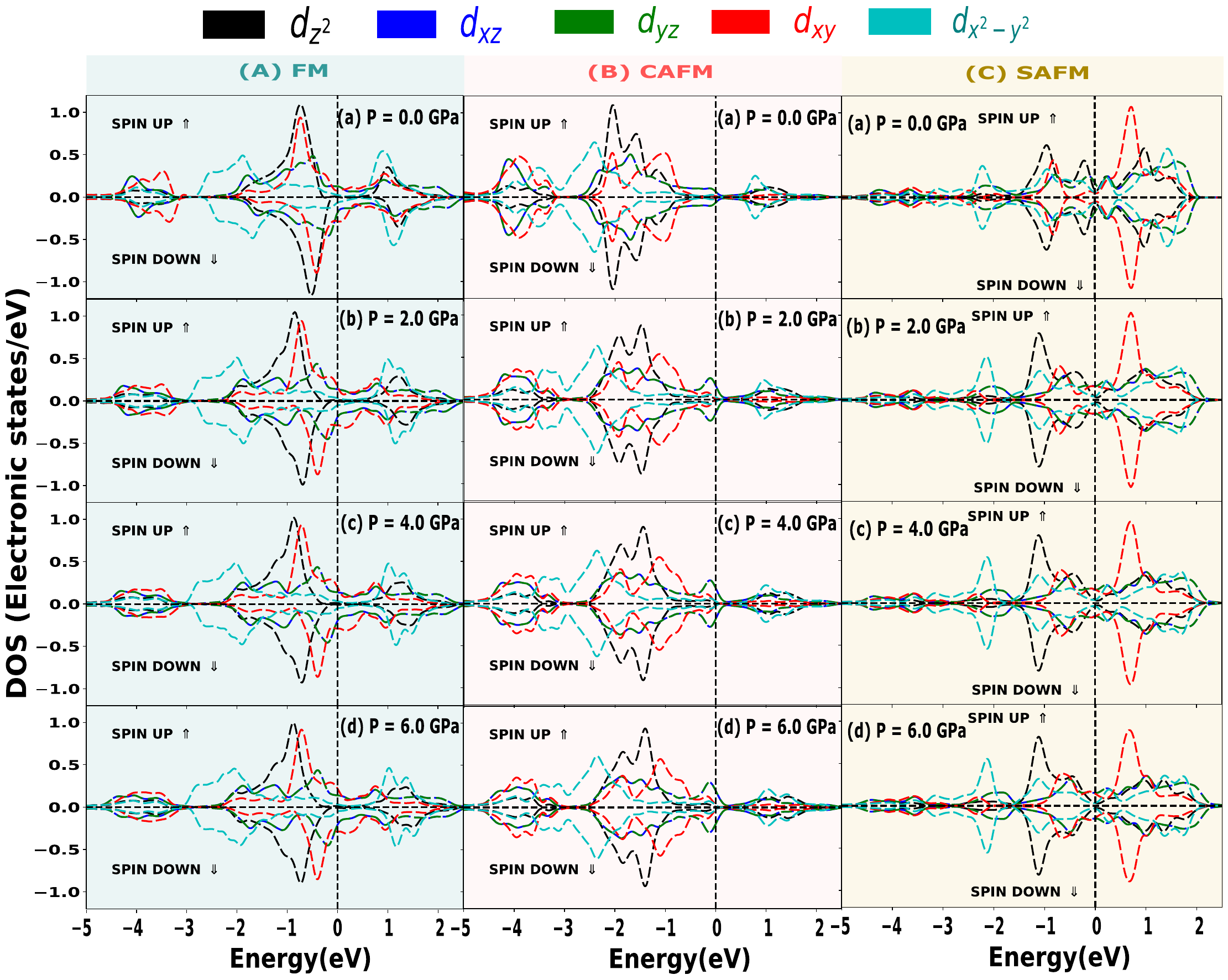}
 \caption{ Variation of spin resolved partial DOS of all the Fe-3d orbitals with pressure in different magnetic phases.}
 \label{pdos}
\end{figure*}
%----------------------------------------------------------

Spin resolved DOS of $t_{2g}$ and $e_g$ orbitals are presented in Fig.\ref{pdos}. Higher partial DOS of $t_{2g}$ in comparison to $e_g$ around the Fermi level is conspicuous in FM and CAFM at ambient pressure. All the five d-orbitals have finite contribution near the Fermi level in case of SAFM. Even $d_{x^2-y^2}/d_{z^2}$ have greater contribution in comparison to $d_{xy}$ around Fermi level. As soon as external pressure is applied on the system, partial DOS of the $e_g$ states are observed to be quenched.